\documentclass[letterpaper]{article}
\usepackage{amsfonts}

\usepackage{graphicx}
\usepackage{amsmath}

\newtheorem{theorem}{Theorem}

\newtheorem{algorithm}[theorem]{Algorithm}

\begin{document}

\title{Control of Switched Networks via Quantum Methods}
\author{Kathryn L.Flores, Viswanath Ramakrishna \\
Department of Mathematical Sciences and Center for Signals,\\
Systems and Telecommunications\\
The University of Texas at Dallas,\\
P.O. Box 830688\\
Richardson, TX 75083\\
e-mail: kflores@utdallas.edu and vish@utdallas.edu\\
Supported in part by the National Science Foundation under grant DMS-0072415.}
\date{}
\maketitle

\begin{abstract}
We illustrate a technique for specifying piecewise constant controls for
classes of switched electrical networks, typically used in converting power
in a dc-dc converter. This procedure makes use of decompositions of $SU(2)$
to obtain controls that are piecewise constant and can be constrained to be
bang-bang with values $0$ or $1$. Complete results are presented for a third
order network first. An example, which shows that the basic strategy is
viable for fourth order circuits, is also given. The former evolves on $%
SO(3) $, while the latter evolves on $SO(4)$. Since the former group is
intimately related to $SU(2)$ while the latter is related to $SU(2)\times
SU(2)$, the methodology of this paper uses factorizations of $SU(2)$. The
systems in this paper are single input systems with drift. In this paper, no
approximations or other artifices are used to remove the drift. Instead, the
drift is important in the determination of the controls. Periodicity
arguments are rarely used. \medskip

\noindent\textbf{Keywords: }bang-bang controls, piecewise constant controls,
Lie group, bilinear system, switched electrical network.
\end{abstract}

\section{\protect\bigskip Introduction}

In this paper the problem of explicit control of a class of switched
electrical, lossless networks is considered. Specifically, it is shown how
to determine explicitly piecewise controls, which can be constrained to take
only the values $1$ or $0$, to achieve state transfers. Complete results are
obtained for a third order lossless network, which has been studied before
in \cite{c}, \cite{d}, \cite{b}. The thesis, \cite{c}, provides the model
and assesses the controllability of the network. The paper, \cite{b}, uses
averaging to provide periodic controls for approximate state preparation.
\textit{The same reference also emphasizes the desirability of finding
bang-bang controls (with values $1$ or $0$), since this mode of control is
closer to physical reality.} In this paper a constructive protocol for
precisely such a bang-bang control is provided. A fourth order network is
also studied and preliminary results on certain explicit state transfers via
bang-bang controls are provided.

The energy conservation of the networks implies that they evolve on $SO(3)$
(respectively $SO(4)$). For the third order network, the problem of
bang-bang controls is susceptible to Euler factorizations (though non-Euler
factorizations are also pertinent). However for constructiveness, explicit
formulae, providing the Euler angles as expressions in the entries of the
target state in $SO(3)$, have to be supplied. To the best of our knowledge
such explicit formulae are missing in the literature, especially when the
two generators of $so(3)$ (the Lie algebra of $SO(3)$), desired in the
factorization, are the ones relevant to the model. \textit{It is worth
emphasizing that the desired state in $SO(3)$ does not already come
specified with its Euler angles.} Rather, it is described by the nine real
entries which constitute this matrix. Similar issues (with the
technicalities compounded) present themselves for the fourth order network.
It is primarily for this reason that the methodology of this paper uses a
passage to an associated system evolving on $SU(2)$ (respectively $%
SU(2)\times SU(2)$). For the system associated to the third order network it
turns out that Euler angles for $SU(2)$, when the two generators are $%
i\sigma _{x}$ and $i\sigma _{y}$, are needed. These are easier than the
corresponding $SO(3)$ angles to calculate because of two reasons: i) first, $%
SU(2)$ matrices are $2\times 2$ (the special unitarity mitigates the fact
that the entries are complex) and thus, the matrix manipulations (which are
inevitable if explicit formulae are required) are easier; and ii) $SU(2)$
matrices admit the following representation (the Cayley-Klein
representation):
\begin{equation}
S=S\left( \alpha ,\zeta ,\mu \right) =\left(
\begin{array}{cc}
e^{i\zeta }\cos \alpha & e^{i\mu }\sin \alpha \\
e^{i\left( \pi -\mu \right) }\sin \alpha & e^{-i\zeta }\cos \alpha
\end{array}
\right) .  \label{5}
\end{equation}
One such representation is nothing more than the entries written in polar
coordinates. The advantage of (\ref{5}) is that the condition $SS^{\ast
}=S^{\ast }S=I,$ $\det \left( S\right) =1$, is already incorporated. In
contrast to $SO(3),$ side conditions need not be stipulated. The attendant
formulae, for even the $SU(2)$ Euler angles, are messier if Cartesian
coordinates were to be used (in our opinion, this is one of the reasons why
explicit formulae for $\left( x,y\right) $ Euler angles for $SO(3)$ are not
available - there is no polar representation for real numbers). Furthermore,
representing the columns of an $SO(3)$ matrix in spherical coordinates is
equally unilluminating. In addition, for finding non-Euler factorizations, $%
SU(2)$ is easier to work with.

The differences between the third order network and the fourth order network
examples are primarily twofold: i) for the fourth order network,
factorizations of $SU(2)$, different from $\sigma _{x},\sigma _{y}$ Euler
angles, are needed. Indeed, the required factorizations are not of the Euler
type.\ Such factorizations are easier to find when working with $SU(2).$ ii)
More importantly, the fourth order network problem amounts to the \textit{%
difficult question of constructive control of two systems with a single
control.} Due to the latter problem our results for the fourth order circuit
are, pending further investigation, applicable under certain conditions on
the circuit. Specifically, the transfers are achieved if any one of a set of
relations between the constants of the circuits are satisfied. In part,
these relations are a by-product of the specific choice of factorizations
used. It should be possible to achieve these relations in practice, since
they are only restrictions on the capacitors and inductors in the circuit.
Work is ongoing to enlarge the class of state transfers and also to
eliminate the restrictions on the constants. These preliminary results are,
to the best of our knowledge, the \textit{first instances }of constructive
controllability for single input systems with drift evolving on $SO(4).$ It
is our opinion that, regardless of the specific model or the control
technique, the most elegant manner to control a system evolving on $SO(4)$
would indeed be to pass to an associated system on $SU(2)\times SU(2)$.
Readers who are skeptical should first attempt to calculate the exponential
of an $so(4)$ matrix without any usage of $SU(2)$ whatsoever. \textit{At a
bare minimum manipulation of }$4\times 4$\textit{\ matrices is required,
whereas passage to }$SU\left( 2\right) \times SU(2)$\textit{\ obviates all
matrix manipulations.} More importantly, finding $e^{A},$ $A\in so(4),$ via
eigenvalues etc., occludes the structure of $A$ in $e^{A}.$ This structure
is relevant to the problem.

Thus, the close relation between $SU(2)$, $SO(3)$ and $SO(4)$ is used for
the network systems. The group, $SU(2)$, plays a prominent role in the
control of many quantum systems (atoms and molecules, Cooper pairs, spin
systems, photons and excitons). This explains the title of the paper. The
rich algebraic structure of the Pauli matrices makes the deduction of the
formulae easier than on the orthogonal groups. However, once a formula has
been found on $SU(2)$ - whether it be for an exponential or bang-bang
controls etc., - it can be transferred easily to the orthogonal group.
\textbf{This is the rationale behind our method.}

Systems such as dc-dc switchmode power converters, in which switched
electrical networks have a significant part, can be implemented in
communication and data handling systems, portable battery-operated equipment
and other applications. Thus, the results of this paper have useful
consequences for these applications. Other strategies for controlling
switched electrical networks use state-space averaging. Leonard and
Krishnaprasad \cite{b} transform these systems into drift free systems and
then apply averaging theory on Lie groups to specify small amplitude,
periodic, open-loop controls for approximate state transfers. The approach
of Sira-Ramirez \cite{g}, based on variable structure systems theory and
sliding regimes, provides feedback controls for switched electrical
networks. In contrast, the method in this paper obtains piecewise constant
controls which further can be taken to be 0 or 1, corresponding to the
position of the switch. From the results of Jurdjevic and Sussmann \cite{h}
it is known that bang-bang controls with values of 0 and 1 can be used to
prepare any target. Thus, the paper provides constructive illustrations of
the work in \cite{h}. \textit{It is emphasized that the approach taken in
this paper does not resort to techniques for driftless systems by either i.)
removing the drift via approximations or other methods which work only in
fortuitous situations or ii.) by making use of periodicity.} Arguments
relying on periodicity are invalid in general \cite{a} and can lead to
expensive controls even when valid. In this paper the only time periodicity
is used is to rewrite free evolution terms with negative drift coefficients
as free evolution terms with positive drift coefficients.

The balance of this paper is organized as follows. In section 2, the
relations between the unitary and orthogonal groups are reviewed. In the
next section, the precise model for the $SO(3)$ network is presented.
Controls for this system are obtained in section 4. This section also
contains the relevant formulae for the desired Euler angles. These are used
to provide, first, piecewise constant controls and then bang-bang controls.
The fifth section provides an illustration of the techniques for the fourth
order network. The final section offers some conclusions.

\section{$SU(2)$, $SO(3)$ and $SO(4)$}

The Lie algebras $su(2)=\{V\in \mathbb{C}^{2\times 2}\mid V^{\ast }=-V\}$
and $so(3)=\{W\in \mathbb{R}^{3\times 3}\mid W^{T}=-W\}$ are isomorphic via
the following explicit isomorphism, \cite{i}:

\begin{equation}
\psi \left[ -\frac{i}{2}\left( a\sigma _{x}+b\sigma _{y}+c\sigma _{z}\right) %
\right] =\left(
\begin{array}{ccc}
0 & -c & b \\
c & 0 & -a \\
-b & a & 0
\end{array}
\right) ,  \label{4}
\end{equation}
with
\begin{equation*}
\sigma _{x}=\left(
\begin{array}{cc}
0 & 1 \\
1 & 0
\end{array}
\right) ,\ \sigma _{y}=\left(
\begin{array}{cc}
0 & -i \\
i & 0
\end{array}
\right) \text{ and }\sigma _{z}=\left(
\begin{array}{cc}
1 & 0 \\
0 & -1
\end{array}
\right) .
\end{equation*}
Similarly, there is a group homomorphism $\phi :SU(2)\rightarrow SO(3),$
\cite{i}$,$ obtained by considering the linear (vector space) map, $%
R_{U}:su(2)\rightarrow su(2)$, which for a fixed $U\in SU(2)$ is given by $%
R_{U}(A)=UAU^{\ast }$. Identifying $su(2)$ with $R^{3}$, it can be shown
that $R_{U}\in SO(3)$. The group homomorphism, $\phi ,$ just associates $U$
to $R_{U}$. Finally, it can be shown, via a direct calculation using the
Rodrigues' formula, that $\phi (e^{K})=e^{\psi (K)}$.

The groups $SO(4)$ and $SU(2)$ are related as follows. First, identifying
the quaternions with $R^{4}$ via $e_{1}=1,\ e_{2}=i,\ e_{3}=j,\ e_{4}=k$,
leads to the following association, $I$, between a pair of unit quaternions,
$p,q$ and a linear map from $R^{4}$ to $R^{4},$ \cite{i}:
\begin{equation*}
I(p,q)=\text{{the linear map}, }x\rightarrow pxq^{-1}.
\end{equation*}
It can be shown that $I(p,q)$ is an element of $SO(4)$. Further, it is well
known, \cite{i}, that the group of unit quaternions is explicitly isomorphic
to $SU(2)$. This then leads to a group homomorphism, $\tilde{\phi}%
:SU(2)\times SU(2)\rightarrow SO(4)$. It can be shown, via a direct
calculation, that there is an associated Lie algebra isomorphism, $\tilde{%
\psi}:su(2)\times su(2)\rightarrow so(4)$, which satisfies $\tilde{\phi}%
(e^{K_{1}},e^{K_{2}})=e^{\tilde{\psi}\left( K_{1}\times K_{2}\right) },$ for
any $K_{1}\times K_{2}$ in $su(2)\times su(2)$. $\tilde{\psi}$ is given by
\begin{equation}
\tilde{\psi}\left( K_{1}\times K_{2}\right) =\left(
\begin{array}{cccc}
0 & -a_{1} & -a_{2} & -a_{3} \\
a_{1} & 0 & -b_{3} & b_{2} \\
a_{2} & b_{3} & 0 & -b_{1} \\
a_{3} & -b_{2} & b_{1} & 0
\end{array}
\right) ,  \label{31}
\end{equation}
where
\begin{equation}
K_{1}=\left(
\begin{array}{cc}
\frac{i}{2}\left( a_{1}+b_{1}\right) & \frac{1}{2}\left[ \left(
a_{2}+b_{2}\right) +i\left( a_{3}+b_{3}\right) \right] \\
\frac{1}{2}\left[ -\left( a_{2}+b_{2}\right) +i\left( a_{3}+b_{3}\right)
\right] & -\frac{i}{2}\left( a_{1}+b_{1}\right)
\end{array}
\right)  \label{32}
\end{equation}
and
\begin{equation}
K_{2}=\left(
\begin{array}{cc}
\frac{i}{2}\left( b_{1}-a_{1}\right) & \frac{1}{2}\left[ \left(
b_{2}-a_{2}\right) +i\left( b_{3}-a_{3}\right) \right] \\
\frac{1}{2}\left[ -\left( b_{2}-a_{2}\right) +i\left( b_{3}-a_{3}\right)
\right] & -\frac{i}{2}\left( b_{1}-a_{1}\right)
\end{array}
\right) .  \label{33}
\end{equation}

Thus, given a system $\dot{V}=\tilde{A}V+\tilde{B}Vu(t),$ $V\in SO(3)$, one
can associate a system, $\dot{U}=AU+BUu(t),$ $U\in SU(2)$, where $A=\psi
^{-1}(\tilde{A})$ and $B=\psi ^{-1}(\tilde{B}),$ to it. Now preparing a
target, $S$ in $SO(3)$ with piecewise constant controls amounts to factoring
$S$ as $\Pi _{k=1}^{Q}e^{(a_{k}\tilde{A}+b_{k}\tilde{B})}$, with $a_{k}>0$,
if $b_{k}\neq 0$. The condition, $b_{k}$ is either $0$ or $b_{k}=a_{k}$, is
equivalent to preparing $S$ with controls only taking values $1$ or $0$. As
mentioned in the introduction, obtaining such factorizations explicitly is
easier for $SU(2)$. Hence, we work with the second system and factorize any
matrix $T$ in $SU(2)$, such that $\phi (T)=S$, as $T=%
\prod_{k=1}^{Q}e^{(a_{k}A+b_{k}B)}$ with either $a_{k}>0$, if $b_{k}\neq 0$
etc. Recapitulating the preparation of a target $S$ in $SO(3)$ by
associating it to a target $T$ in $SU(2),$
\begin{equation}
S=\phi \left( T\right) =\phi \left( \prod_{k=1}^{Q}e^{a_{k}A+b_{k}B}\right) ,
\label{3}
\end{equation}
because $\phi $ is a homomorphism this gives
\begin{equation}
S=\prod_{k=1}^{Q}\phi \left( e^{a_{k}A+b_{k}B}\right) ,
\end{equation}
and since $\phi (e^{K})=e^{\psi (K)},$%
\begin{equation}
S=\prod_{k=1}^{Q}e^{a_{k}\tilde{A}+b_{k}\tilde{B}}.  \label{13}
\end{equation}
\textit{Therefore the same controls that prepare }$T$\textit{\ also prepare }%
$S.$ The corresponding control values are, of course, $\frac{b_{k}}{a_{k}}$.

Likewise, given a system $\dot{V}=\tilde{A}V+\tilde{B}Vu(t),$ $V\in SO(4)$,
\textbf{two systems controlled by a single control }$\mathbf{u}\left(
t\right) $, are associated to it via, $\dot{U_{1}}%
=A_{1}U_{1}+B_{1}U_{1}u(t), $ $U_{1}\in SU(2)$ and $\dot{U_{2}}%
=A_{2}U_{2}+B_{2}U_{2}u(t),$ $U_{2}\in SU(2)$. Here, $(A_{1},A_{2})=\tilde{%
\psi}^{-1}(\tilde{A})$ and $(B_{1},B_{2})=\tilde{\psi}^{-1}(\tilde{B})$.
Given a target, $S\in SO(4)$, we prepare any $(T_{1},T_{2})$ such that $%
\tilde{\phi}(T_{1},T_{2})=S$. Usage of piecewise constant controls means
that both the $T_{i}$ have to be factorized as $T_{i}=\Pi
_{k=1}^{Q}e^{(a_{k}A_{i}+b_{k}B_{i})}$, with the \textit{same} $Q$ and
\textit{same} $a_{k}$ and $b_{k}$ for all $k=1,\ldots ,Q$. The usual
stipulations, $a_{k}>0$ if $b_{k}\neq 0$ etc., apply here too.\bigskip

\noindent \textbf{Remark 1:} It is well known, \cite{i}, that the kernel of $%
\phi $ is $\left\{ +I_{2},-I_{2}\right\} $ and that the kernel of $\tilde{%
\phi}$ is $\left\{ \left( I_{2},I_{2}\right) ,\left( -I_{2},-I_{2}\right)
\right\} .$ In this paper we do not make systematic use of this extra degree
of freedom.

\section{The Third Order Network}

\begin{figure}[t]
 \centering
 \includegraphics[scale=0.7]{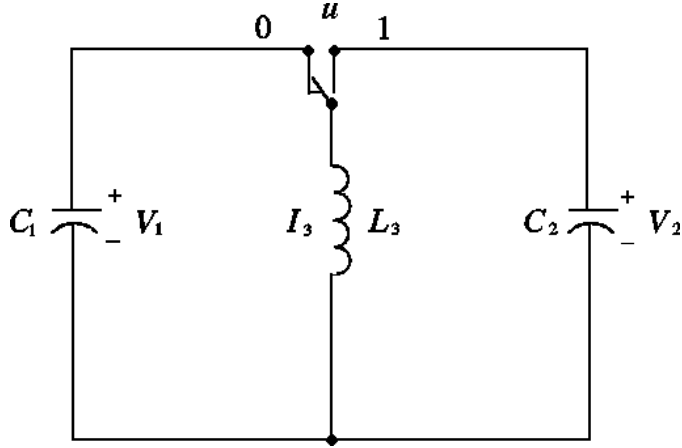}
 \caption{Third Order Switched Network}
 \label{dd}
\end{figure}

In this section, a switched electrical network with three circuit elements
and no external constant power sources is considered. The switched
electrical network examined here is identical to the one used by Leonard,
Krishnaprasad and Wood \cite{b}, \cite{c}. This network consists of two
capacitors $C_{1}$ and $C_{2}$ with corresponding voltages $V_{1}$ and $%
V_{2} $ (see Figure \ref{dd}). These capacitors are connected by a
switch and an inductor $L_{3}$ with current $I_{3}$. The position
of the switch connected to a control takes only the values of $0$
or $1.$ The control objective is to transfer energy from $C_{1}$
to $C_{2}$ by means of the inductor. Such systems, in the absence
of external loads, can be modeled \cite{b} by defining the network
state vector $\mathbf{x}=\left(
x_{1},x_{2},x_{3}\right) ^{T}$ as $x_{1}=\sqrt{C_{1}}V_{1}$, $x_{2}=\sqrt{%
C_{2}}V_{2}$, and $x_{3}=\sqrt{L_{3}}I_{3}$. Let $\omega _{1}=1/\sqrt{%
C_{1}L_{3}}$ and $\omega _{2}=1/\sqrt{C_{2}L_{3}}$. Then the system is
\begin{equation}
\frac{d}{dt}\mathbf{x}=\left(
\begin{array}{ccc}
0 & 0 & \omega _{1}\left( 1-u\right) \\
0 & 0 & \omega _{2}u \\
-\omega _{1}\left( 1-u\right) & -\omega _{2}u & 0
\end{array}
\right) x=\left( \tilde{A}+\tilde{B}u\right) \mathbf{x.}  \label{1}
\end{equation}

If the control takes a constant value $%
u$ for a time $t$, the state of the system can be written as
\begin{equation}
\mathbf{x}\left( t\right) =e^{\left( \tilde{A}+\tilde{B}u\right) t}\mathbf{x}%
\left( 0\right) .
\end{equation}

The system on $SU\left( 2\right) $ associated to the system on $SO\left(
3\right) ,$ from section 2, is
\begin{equation}
\frac{d}{dt}U=\left( -\frac{i}{2}\omega _{1}\sigma _{y}+\frac{i}{2}\left(
\omega _{1}\sigma _{y}+\omega _{2}\sigma _{x}\right) u\right) U\mathbf{.}
\label{9}
\end{equation}

\section{Quantum Control Techniques}

Preparing the final state, $\mathbf{x}_{f},$ is equivalent to the
preparation of one of an infinite family of $SO(3)$ matrices such that $%
\mathbf{x}_{f}=S\mathbf{x}\left( 0\right) .$ This, in turn, defines a family
of targets in $SU(2)$ that can be associated with each such $S.$ Each target
$S\in SO(3)$ corresponds to two targets in $SU(2).$ Preparing either of
these two targets amounts to preparing $S.$

A target $T\in $ $SU(2),$ can be written as
\begin{equation}
T=\exp \left[ -\frac{i}{2}\left( a\sigma _{x}+b\sigma _{y}+c\sigma
_{z}\right) \right] ,  \label{14}
\end{equation}
where $a,b,c\in \mathbb{R}.$ Let $\lambda =\frac{1}{2}\sqrt{a^{2}+b^{2}+c^{2}%
}$, $s=-\frac{i}{2}\left( a\sigma _{x}+b\sigma _{y}+c\sigma _{z}\right) $
and $\mathbf{p}=\left(
\begin{array}{ccc}
a & b & c
\end{array}
\right) ^{T}.$ An expression for the Lie group homomorphism, $\phi ,$
described in section 2 is obtained from Rodrigues' formula \cite{f},
\begin{equation}
\exp \left[ \psi \left( s\right) \right] =I\cos 2\lambda +\frac{\sin
2\lambda }{2\lambda }\psi \left( s\right) +\frac{1-\cos 2\lambda }{4\lambda
^{2}}\mathbf{pp}^{T},  \label{16}
\end{equation}
by using equation (\ref{4}) and the fact that $\phi \left( T\right) =\exp
\left[ \psi \left( s\right) \right] .$ Now the problem of finding controls
to drive the switched network on $SO(3)$ from $\mathbf{x}\left( 0\right) $
to $\mathbf{x}_{f}$ is converted to finding controls that prepare $T\in
SU(2).$ This is accomplished by using an appropriate decomposition of $T.$

\subsection{Decompositions of a target in SU(2)}

The general problem of preparing targets in $SU(2)$ was considered in \cite
{a}. In that work the theory requires that the drift and control matrices $A$
and $B$ be orthonormal. Orthonormality of $A$ and $B$ can be achieved by
preliminary controls. The use of preliminary controls precludes the
construction of bang-bang controls. Since preliminary orthonormalization of
the matrices $A$ and $B$ is not used here, the results do not follow
directly from \cite{a}. However, the general framework of that paper is
helpful in this work.

Consider the general problem of preparing a target for the system (\ref{9})
in $SU(2)$. The decompositions of elements of $SU(2)$ considered in this
paper are based on the fact that $A$ and $B$ in (\ref{9}) are linear
combinations of $i\sigma _{x}$ and $i\sigma _{y}.$ By writing the entries of
$T$ in the Cayley-Klein representation (\ref{5}), various decompositions can
be obtained. We describe three of these factorizations, one of which is used
for general piecewise constant controls and the other two for bang-bang
controls. First as shown in \cite{a}, matrices in $SU(2)$ may be decomposed
into the following form:
\begin{equation}
T\left( \alpha ,\zeta ,\mu \right) =e^{ip\sigma _{z}}V\left( \gamma \right)
e^{i\left( \zeta -p\right) \sigma _{z}},  \label{8}
\end{equation}
for any $p\in \mathbb{R}$ and
\begin{equation}
V\left( \gamma \right) =\exp \left(
\begin{array}{cc}
0 & i\gamma \\
i\bar{\gamma} & 0
\end{array}
\right) =\exp \left[ \left( -\textnormal{Im}\gamma \right) i\sigma
_{y}+\left( \textnormal{Re}\gamma \right) i\sigma _{x}\right]
\label{17}
\end{equation}
for $\gamma \in \mathbb{C}.$ In equation (\ref{8}), $\gamma =\alpha \exp
i\left( \zeta +\mu -2p-\frac{\pi }{2}\right) $ and $p$ can be chosen so that
$V\left( \gamma \right) $ is a free evolution factor. For the switched
electrical network considered in this paper, the first and third factors of
equation (\ref{8}) have a useful decomposition. In \cite{e}, it was proved
that the exponential of the third Pauli matrix can be expressed as a product
of two factors
\begin{equation}
e^{iL\sigma _{z}}=V\left( \gamma _{1}\right) V\left( \gamma _{2}\right) ,
\label{6}
\end{equation}
where $L\in \mathbb{R}$. Thus it follows from equation (\ref{8}) that $%
T=\prod_{k=1}^{Q}V\left( \gamma _{k}\right) ,$ where $1\leq Q\leq 5.$ In (%
\ref{6}), let $\gamma _{k}=\frac{\pi }{2}e^{i\theta _{k}},$ $k=1,2,$ then it
holds that $L=\theta _{1}-\theta _{2}+\pi .$ Since $L$ can be taken as an
element of $[0,2\pi ),$ it follows that $\left| L-\pi \right| <\pi ,\ L\neq
0.$ Thus for $L\neq 0,$ $\gamma _{1}$ and $\gamma _{2}$ can be chosen so
that $\left| \theta _{1}-\theta _{2}\right| <\pi .$ This means that $\gamma
_{1}$ and $\gamma _{2}$ may be selected to lie any open half-plane. In other
words, one can ensure that $a_{k}>0,$ $\forall k,$ as
\begin{equation}
V\left( \gamma _{k}\right) =\exp \left[ b_{k}\frac{\omega _{2}}{2}i\sigma
_{x}+\left( b_{k}-a_{k}\right) \frac{\omega _{1}}{2}i\sigma _{y}\right] .
\label{18}
\end{equation}
It follows that the half-plane of interest is $\textnormal{Im}\gamma _{k}>-\frac{%
\omega _{1}}{\omega _{2}}\textnormal{Re}\gamma _{k}.$ This
decomposition of the target in $SU(2)$ provides piecewise constant
controls with no further restrictions.\bigskip

\noindent \textbf{Remark 2:} The utility of the decomposition (\ref{6}) is
the following. Together with equation (\ref{8}) it provides a factorization
of any $T$ in $SU\left( 2\right) $ of the type given by equation (\ref{3}),
with generally lower values of $\sqrt{a_{k}^{2}+b_{k}^{2}}$ for each $k,$
than would a factorization provided by the Euler parametrization (eqtns (\ref
{21})-(\ref{24}) below). This can be seen by viewing each factor in both of
these decompositions as a matrix, $V\left( \gamma \right) $ (cf. eqtn (\ref
{17})).

The complex numbers, $\gamma ,$ in the factorization provided by equation (%
\ref{6}) each have a radial coordinate at most $\frac{\pi }{2},$ whereas the
radial coordinates due to Euler factorizations could be as high as $2\pi .$
This causes the former factorization to yield, generally, lower values for
(individual and cumulative) $\sqrt{a_{k}^{2}+b_{k}^{2}}.$ Since $a_{k}$
represents the duration and $b_{k}$ the power (=duration$\times $amplitude)
of the $k$th pulse, this suggests that equation\ (\ref{6}) is preferable for
the simultaneous minimization of these two competing constraints, as long as
it is reasonable to use any piecewise constant control.\bigskip \bigskip

Next consider bang-bang controls. For free evolution, $u_{k}=0,$ $b_{k}=0$
and from equation (\ref{18}), $V\left( \gamma _{k}\right) =\exp \left( -a_{k}%
\frac{\omega _{1}}{2}i\sigma _{y}\right) ,$ or in other words, $\theta _{k}=%
\frac{\pi }{2}.$ When $u_{k}=1,$ this means that $a_{k}=b_{k},$ so that $%
V\left( \gamma _{k}\right) =\exp \left( b_{k}\frac{\omega _{2}}{2}i\sigma
_{x}\right) $ which corresponds to the phase $\theta _{k}=0.$ These facts
lead to the consideration of the following decomposition of the exponential
of the third Pauli matrix:
\begin{equation}
e^{iL\sigma _{z}}=e^{-i\frac{7\pi }{4}\sigma _{y}}e^{iL\sigma _{x}}e^{-i%
\frac{\pi }{4}\sigma _{y}}.  \label{7}
\end{equation}
The first and third factors of equation (\ref{7}) are free evolution and the
second factor is obtained with a control pulse of $1.$ In each factor of\ (%
\ref{7}) the drift coefficient is a positive number. With this decomposition
the target $T$ from equation (\ref{8}), with $p$ chosen so that $V\left(
\gamma \right) $ is a free evolution factor, is prepared by at most seven
factors of which at most two are control pulses. Now we consider another
decomposition of $T$ from which $T$ is prepared by at most three factors.

A different bang-bang protocol is obtained by the following. It is shown in
\cite{q} that the target $T$ can be expressed as
\begin{equation}
T\left( \alpha ,\zeta ,\mu \right) =e^{iD\sigma _{x}}e^{iE\sigma
_{y}}e^{iF\sigma _{x}},  \label{21}
\end{equation}
where $D$, $E$ and $F$ are solutions to the relations
\begin{eqnarray}
\cos (E) &=&\pm \sqrt{\cos ^{2}\zeta \cos ^{2}\alpha +\sin ^{2}\mu \sin
^{2}\alpha }  \label{22} \\
\sin (D-F) &=&\pm \frac{\sin \zeta \cos \alpha }{\sqrt{\sin ^{2}\zeta \cos
^{2}\alpha +\cos ^{2}\mu \sin ^{2}\alpha }}  \label{23} \\
\sin (D+F) &=&\pm \frac{\sin \mu \sin \alpha }{\sqrt{\cos ^{2}\zeta \cos
^{2}\alpha +\sin ^{2}\mu \sin ^{2}\alpha }}.  \label{24}
\end{eqnarray}

\noindent \textbf{Remark 3:} The expressions for the Euler angles given by
equations (\ref{22})-(\ref{24}) were obtained by an explicit matrix
calculation. Indeed, it is known that an Euler angle factorization, with
factors that are exponentials of $i\sigma _{z}$ and $i\sigma _{y},$ exists
with a maximum of three factors. The orthogonality of the pairs $\left(
i\sigma _{z},i\sigma _{y}\right) $ and $\left( i\sigma _{x},i\sigma
_{y}\right) $ suggests an obvious Lie algebra isomorphism of $su\left(
2\right) $ with itself. This suggests that it should be possible to find a
factorization of the type in equation (\ref{21}). This matrix calculation is
facilitated by an explicit expression for the exponential of an $su\left(
2\right) $ matrix (which, incidently, is easier to manipulate than the
corresponding $so\left( 3\right) $ expression). Even though there is \ a
natural geometric equivalence between the pairs of generators of $su\left(
2\right) ,$ $\left( i\sigma _{z},i\sigma _{y}\right) $ and $\left( i\sigma
_{x},i\sigma _{y}\right) ,$ the expressions for the Euler angles are not
simple consequences of one another. It is routine to show that the $\left(
i\sigma _{z},i\sigma _{y}\right) $ Euler angles are linear in the
Cayley-Klein coordinates, whereas, equations (\ref{22})-(\ref{24})
demonstrate that the $\left( i\sigma _{x},i\sigma _{y}\right) $ Euler angles
involve transcendental functions. \bigskip \bigskip

In the decomposition of $T$ \ in equation (\ref{21}) the second factor is
free evolution and control pulses of $1$ are used to obtain the first and
third factors. This decomposition of $T$ prepares the target with at most
three factors with no more than two control pulses. In summary, we have:

\begin{algorithm}
\label{A1}Piecewise constant controls.
\end{algorithm}

\begin{enumerate}
\item  Given an initial state $\mathbf{x}(0)$ and a final state $\mathbf{x}%
_{f},$ choose the numbers $a,b,$ and $c$ in equation (\ref{16}) so that the
target on $SO(3),$ $S=\exp \left[ \psi \left( s\right) \right] ,$ satisfies $%
\mathbf{x}_{f}=S\mathbf{x}\left( 0\right) .$

\item  Write the entries of $S$ in polar coordinates as in equation (\ref{5}%
).

\item  Choose $p$ in equation (\ref{8}) so that $V\left( \gamma \right) $ is
a free evolution factor.

\item  For each of the first and third factors of equation (\ref{8}) use the
decomposition in equation (\ref{6}). Select the phases of $\gamma _{1}=\frac{%
\pi }{2}e^{i\theta _{1}}$ and $\gamma _{2}=\frac{\pi }{2}e^{i\theta _{2}}$
in equation (\ref{6}) so that $L=\theta _{1}-\theta _{2}+\pi $ and the drift
coefficients are positive numbers.
\end{enumerate}

\begin{algorithm}
\label{A2}Bang-bang controls I
\end{algorithm}

Steps one through three are the same as for piecewise constant controls.

Step four: For each of the first and third factors of equation (\ref{8}) use
the decomposition in equation (\ref{7}).

\begin{algorithm}
\label{A3}Bang-bang controls II
\end{algorithm}

Steps one and two are the same as for piecewise constant controls.

Step three: Solve for $D$, $E$ and $F$ in the decomposition of $T$ given by
equation (\ref{21}).

\subsection{Example}

For the switched network in Figure \ref{dd}, if $C_{1}=0.1,$ $C_{2}=0.2$ and
$L_{3}=0.5,$ then $A=-\sqrt{5}i\sigma _{y}$ and $B=\frac{i}{2}\left( 2\sqrt{5%
}\sigma _{y}+\sqrt{10}\sigma _{x}\right) .$ Suppose the initial state vector
$\mathbf{x}(0)=(1,0,0)^{T}$ and the final state vector $\mathbf{x}%
_{f}=(0,-1,0)^{T}.$ Intermediate points for the system to traverse may be
specified such as $\mathbf{x}(t1)=(1/\sqrt{2},0,-1/\sqrt{2})^{T}$ for the
first intermediate point\ and $\mathbf{x}(t2)=(0,-1/\sqrt{2},-1/\sqrt{2}%
)^{T} $ for the second intermediate point \cite{b}. Suppose that it is
desired for the system to pass through the intermediate points $\mathbf{x}%
\left( t1\right) $ and $\mathbf{x}(t2).$ Then three targets $T_{1}$, $T_{2}$
and $T_{3}$ in $SU(2)$ must be prepared so that $\mathbf{x}\left( t1\right)
=\phi \left( T_{1}\right) \mathbf{x}(0)$, $\mathbf{x}(t2)=\phi \left(
T_{2}\right) \mathbf{x}(t1)$ and $\mathbf{x}_{f}=\phi \left( T_{3}\right)
\mathbf{x}(t2).$

The target $T_{1}$ is determined by the the Lie group homomorphism
\begin{equation}
\phi \left( T\right) =\left(
\begin{array}{lll}
\cos 2\lambda +\frac{a^{2}\left( 1-\cos 2\lambda \right) }{4\lambda ^{2}} &
\frac{ab\left( 1-\cos 2\lambda \right) }{4\lambda ^{2}}-\frac{c\sin 2\lambda
}{2\lambda } & \frac{ac\left( 1-\cos 2\lambda \right) }{4\lambda ^{2}}+\frac{%
b\sin 2\lambda }{2\lambda } \\
\frac{ab\left( 1-\cos 2\lambda \right) }{4\lambda ^{2}}+\frac{c\sin 2\lambda
}{2\lambda } & \cos 2\lambda +\frac{b^{2}\left( 1-\cos 2\lambda \right) }{%
4\lambda ^{2}} & \frac{bc\left( 1-\cos 2\lambda \right) }{4\lambda ^{2}}-%
\frac{a\sin 2\lambda }{2\lambda } \\
\frac{ac\left( 1-\cos 2\lambda \right) }{4\lambda ^{2}}-\frac{b\sin 2\lambda
}{2\lambda } & \frac{bc\left( 1-\cos 2\lambda \right) }{4\lambda ^{2}}+\frac{%
a\sin 2\lambda }{2\lambda } & \cos 2\lambda +\frac{c^{2}\left( 1-\cos
2\lambda \right) }{4\lambda ^{2}}
\end{array}
\right) .  \label{12}
\end{equation}
and $\mathbf{x}\left( t1\right) =\phi \left( T_{1}\right) \mathbf{x}(0)$
which lead to the relation
\begin{equation}
\left(
\begin{array}{l}
\cos 2\lambda _{1}+\frac{a_{1}^{2}\left( 1-\cos 2\lambda _{1}\right) }{%
4\lambda _{1}^{2}} \\
\frac{a_{1}b_{1}\left( 1-\cos 2\lambda _{1}\right) }{4\lambda _{1}^{2}}+%
\frac{c_{1}\sin 2\lambda _{1}}{2\lambda _{1}} \\
\frac{a_{1}c_{1}\left( 1-\cos 2\lambda _{1}\right) }{4\lambda _{1}^{2}}-%
\frac{b_{1}\sin 2\lambda _{1}}{2\lambda _{1}}
\end{array}
\right) =\left(
\begin{array}{c}
\frac{1}{\sqrt{2}} \\
0 \\
-\frac{1}{\sqrt{2}}
\end{array}
\right) .  \label{19}
\end{equation}
Also, $\phi \left( T_{1}\right) \in SO(3)$ requires that $\det \left[ \phi
\left( T_{1}\right) \right] =1.$ Let $a_{1}=0$ and $c_{1}=0,$ then $2\lambda
_{1}=\left| b_{1}\right| $ and $\cos 2\lambda _{1}=\frac{1}{\sqrt{2}}.$
Setting $b_{1}=\frac{\pi }{4}$ gives
\begin{equation}
\phi \left( T_{1}\right) =\left(
\begin{array}{ccc}
\frac{1}{\sqrt{2}} & 0 & \frac{1}{\sqrt{2}} \\
0 & 1 & 0 \\
-\frac{1}{\sqrt{2}} & 0 & \frac{1}{\sqrt{2}}
\end{array}
\right) .
\end{equation}
Therefore, $T_{1}=\exp \left( -\frac{\pi }{8}i\sigma _{y}\right) $ satisfies
equation (\ref{19}) and $\det \left[ \phi \left( T_{1}\right) \right] =1.$

Similarly, the choice of the target $T_{2}$ must satisfy $\mathbf{x}%
(t2)=\phi \left( T_{2}\right) \mathbf{x}(t1)$ and $\phi \left( T_{2}\right)
\in SO(3).$ By letting $a_{2}=0$ and $b_{2}=0,$ then $2\lambda _{2}=\left|
c_{2}\right| $ and $\sin 2\lambda _{2}=\pm 1.$ Choosing $c_{2}=-\frac{\pi }{2%
}$ leads to
\begin{equation}
\phi \left( T_{2}\right) =\left(
\begin{array}{ccc}
0 & 1 & 0 \\
-1 & 0 & 0 \\
0 & 0 & 1
\end{array}
\right) .
\end{equation}
Hence, $T_{2}=\exp \left( \frac{\pi }{4}i\sigma _{z}\right) $ is a suitable
target in $SU(2).$

A target $T_{3}$ that meets the requirements $\mathbf{x}_{f}=\phi \left(
T_{3}\right) \mathbf{x}(t2)$ and $\phi \left( T_{3}\right) \in SO(3)$ is
obtained by letting $b_{3}=0$ and $c_{3}=0$ from which it follows that $%
2\lambda _{3}=\left| a_{3}\right| $ and $-\frac{1}{\sqrt{2}}\cos a_{3}+\frac{%
1}{\sqrt{2}}\sin a_{3}=-1.$ Set $a_{3}=-\frac{\pi }{4}$ and this gives
\begin{equation}
\phi \left( T_{3}\right) =\left(
\begin{array}{ccc}
1 & 0 & 0 \\
0 & \frac{1}{\sqrt{2}} & \frac{1}{\sqrt{2}} \\
0 & -\frac{1}{\sqrt{2}} & \frac{1}{\sqrt{2}}
\end{array}
\right) .
\end{equation}
So $T_{3}=\exp \left( \frac{\pi }{8}i\sigma _{x}\right) $ is a suitable
target.

\subsubsection{Piecewise constant controls}

Algorithm \ref{A1} is applied to the preparation of $T_{1},$ $T_{2}$ and $%
T_{3}.$ By equations (\ref{17}) and (\ref{18}), the real and imaginary parts
of $\gamma _{k},$ $k=1,\ldots ,Q$ are linear combinations of $a_{k}$ and $%
b_{k}.$ For the network example, each of the factors of $T$ in equation (\ref
{3}) can be represented by
\begin{equation}
\exp \left[ b_{k}\sqrt{\frac{5}{2}}i\sigma _{x}+\left( b_{k}-a_{k}\right)
\sqrt{5}i\sigma _{y}\right]  \label{15}
\end{equation}
for $k=1,\ldots ,Q.$ From this relation and $a_{k}>0$, we find that for each
$k,$ $\textnormal{Im}\gamma _{k}>-\sqrt{2}\textnormal{Re}\gamma _{k}.$ So our choice of $%
\gamma _{k}$ must be above the line $\textnormal{Im}\gamma _{k}=-\sqrt{2}\textnormal{Re}%
\gamma _{k}.$ This means that $\theta _{k}\in \left( -0.3041\pi ,0.6959\pi
\right) ,$ which is in keeping with the fact that $\theta _{k}=\frac{\pi }{2}
$ corresponds to free evolution and $\theta _{k}=0$ when $u_{k}=1$.

For the preparation of $T_{1}=\exp \left( -\frac{\pi }{8}i\sigma _{y}\right)
,$ note that it can be achieved by free evolution with $t1=\frac{\pi }{8%
\sqrt{5}}.$

\begin{figure}[t]
 \centering
 \includegraphics[scale=0.5]{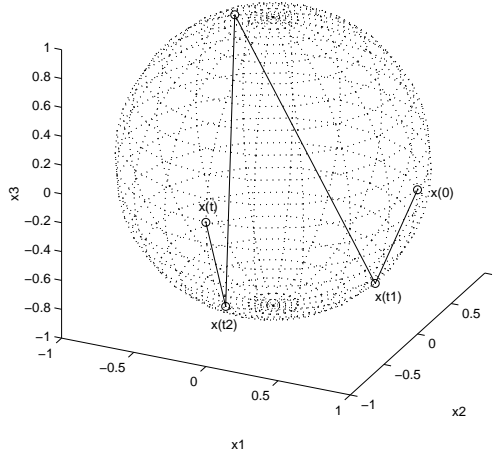}
 \caption{States obtained via general piecewise constant controls}
 \label{bb}
\end{figure}

Now consider preparing $T_{2}=\exp \left( \frac{\pi }{4}i\sigma _{z}\right)
. $ It follows from equation (\ref{6}) that
\begin{equation}
T_{2}=V\left( \gamma _{1}\right) V\left( \gamma _{2}\right) ,
\end{equation}
where $\gamma _{k}=\frac{\pi }{2}e^{i\theta _{k}},$ $k=1,2,$ and the phases
of $\gamma _{1}$ and $\gamma _{2}$ must satisfy $\theta _{2}-\theta _{1}=%
\frac{3\pi }{4}.$ Choose $\theta _{1}=-\frac{\pi }{8}$ and $\theta _{2}=%
\frac{5\pi }{8}.$ Using equation (\ref{15}), the following coefficients are
obtained for $T_{2}$:
\begin{equation}
\begin{tabular}{c|c|c}
$k$ & $a_{k}$ & $b_{k}$ \\ \hline
$1$ & $0.649$ & $0.917$ \\
$2$ & $0.269$ & $-0.380$%
\end{tabular}
.
\end{equation}
These coefficients, when placed in equation (\ref{13}), indeed drive the
system in $SO(3)$ from $\mathbf{x}\left( t1\right) $ to $\mathbf{x}\left(
t2\right) $ as shown in Figure \ref{bb}.

To prepare $T_{3}=\exp \left( \frac{\pi }{8}i\sigma _{x}\right) ,$ observe
that it can be obtained with a control pulse of one. Thus, $%
t_{f}-t2=a_{1}=b_{1}=\frac{\pi }{4\sqrt{10}}.$

In Figure \ref{bb} the lines indicate that the network system is taken from
a point in $\mathbb{R}^{3}$ to another along some undetermined path on the
unit sphere. Because $b_{2}$ is a negative number, $u_{2}$ is negative and,
thus, $u_{2}$ is not a control pulse that represents the position of the
switch.

\subsubsection{Bang-bang controls I}

Now algorithm \ref{A2} is applied to the preparation of $T_{1},$ $T_{2}$ and
$T_{3}.$ As stated previously, the target $T_{1}$ can be prepared by free
evolution and the target $T_{3}$ can be obtained with a control pulse of $1.$
Therefore, it remains to get bang-bang controls for $T_{2}$ so that the
controls represent the position of the switch. It follows from equation (\ref
{7}) that
\begin{equation}
T_{2}=\exp \left( i\frac{\pi }{4}\sigma _{z}\right) =e^{-i\frac{7\pi }{4}%
\sigma _{y}}e^{i\frac{\pi }{4}\sigma _{x}}e^{-i\frac{\pi }{4}\sigma _{y}}.
\label{20}
\end{equation}
Each of the three factors of equation (\ref{20}) are of the form of (\ref{15}%
), so that we have the following coefficients for $T_{2}$:
\begin{equation}
\begin{tabular}{c|c|c}
$k$ & $a_{k}$ & $b_{k}$ \\ \hline
$1$ & $\frac{\pi \sqrt{5}}{20}$ & $0$ \\
$2$ & $\frac{\pi \sqrt{10}}{20}$ & $\frac{\pi \sqrt{10}}{20}$ \\
$3$ & $\frac{7\pi \sqrt{5}}{20}$ & $0$%
\end{tabular}
.
\end{equation}
\begin{figure}[t]
 \centering
 \includegraphics[scale=0.5]{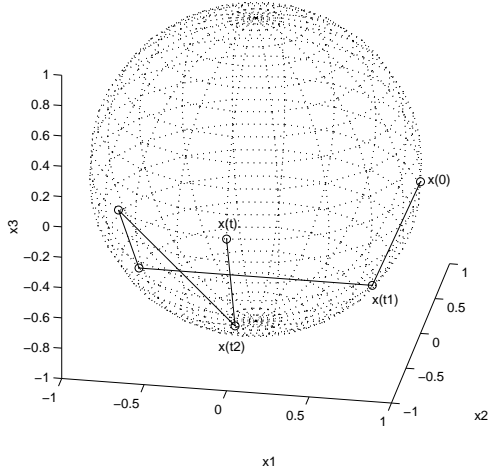}
 \caption{States obtained with bang-bang controls from algorithm \ref{A2}}
 \label{aa}
\end{figure}
These coefficients represent bang-bang controls that drive the system as
shown in Figure \ref{aa}.

As in Figure \ref{bb}, each line in Figure \ref{aa} represents the
system being taken from one point in $\mathbb{R}^{3}$ to another along an
undetermined path on the unit sphere. Controls having values of $0$ and $1$
to prepare $T_{2}$ can also be obtained by utilizing algorithm \ref{A3}.

\subsubsection{Bang-bang controls II}

The preparation of $T_{1},$ $T_{2}$ and $T_{3}$ is now achieved by use of
algorithm \ref{A3} to obtain controls of $0$ and $1$. Again, the target $%
T_{1}$ can be prepared by free evolution and the target $T_{3}$ can be
obtained with a control pulse of $1.$ It remains to prepare $T_{2}=\exp
\left( i\frac{\pi }{4}\sigma _{z}\right) $ by solving for $D$, $E$ and $F$
in the decomposition of $T$ given by equation(\ref{21}). Equation (\ref{5})
is used to find values for $\alpha $, $\zeta $ and $\mu .$ Since
\begin{equation}
T_{2}=\left(
\begin{array}{cc}
\exp \left( i\frac{\pi }{4}\right) & 0 \\
0 & \exp \left( -i\frac{\pi }{4}\right)
\end{array}
\right) ,
\end{equation}
$\alpha =0,$ $\zeta =\frac{\pi }{4}$ and $\mu \in \left[ 0,2\pi \right) .$
From equations (\ref{22}), (\ref{23}) and (\ref{24}) we can choose $D=\frac{%
3\pi }{4}$, $E=-\frac{7\pi }{4}$ and $F=\frac{\pi }{4}.$ These values of $D$%
, $E$ and $F$ correspond to the following coefficients of $T_{2}$:
\begin{equation}
\begin{tabular}{c|c|c}
$k$ & $a_{k}$ & $b_{k}$ \\ \hline
$1$ & $\frac{3\sqrt{10}\pi }{20}$ & $\frac{3\sqrt{10}\pi }{20}$ \\
$2$ & $\frac{7\sqrt{5}\pi }{20}$ & $0$ \\
$3$ & $\frac{\sqrt{10}\pi }{20}$ & $\frac{\sqrt{10}\pi }{20}$%
\end{tabular}
.
\end{equation}
\begin{figure}[t]
 \centering
 \includegraphics[scale=0.5]{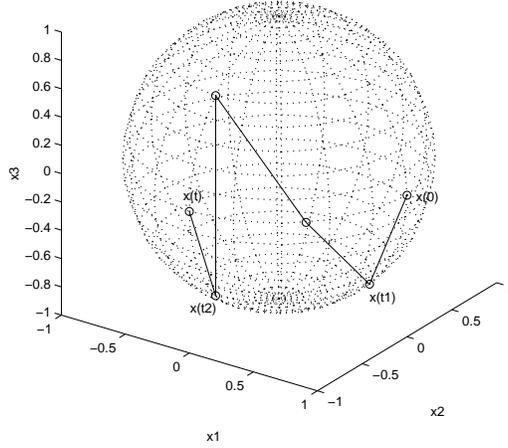}
 \caption{States obtained with bang-bang controls from algorithm \ref{A3}}
 \label{ff}
\end{figure}

The switched network system in $SO(3)$ is driven from
$\mathbf{x}\left( t1\right) $ to $\mathbf{x}\left( t2\right) $
with these coefficients as shown in Figure \ref{ff}. Each line in
Figure \ref{ff} represents the system being taken from one point
in $\mathbb{R}^{3}$ to another along an undetermined path on the
unit sphere.

\section{A Fourth Order Network}

\begin{figure}
 \centering
 \includegraphics[scale=0.7]{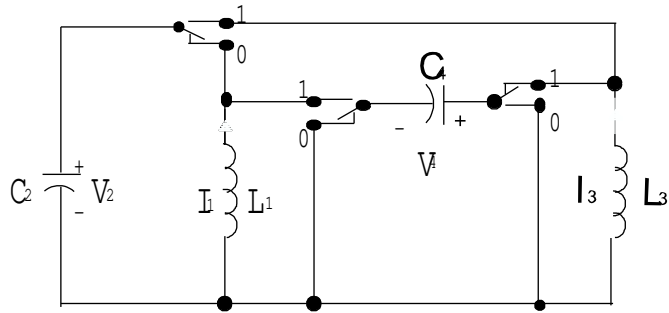}
 \caption{Fourth Order Switched Network}
 \label{FON}
\end{figure}

In this section a fourth order network (see Figure \ref{FON})
taken from \cite{c} is considered and it is shown how to effect
certain state transfers via bang-bang controls. To the best of our
knowledge these state transfers are the first instances of
explicit exact control of systems with drift on $S^{3},$ the
sphere in $\mathbb{R}^{4}.$ The system's equations are:
\begin{equation}
\dot{x}=\left(
\begin{array}{cccc}
0 & -\nu & 0 & 0 \\
\nu & 0 & 0 & 0 \\
0 & 0 & 0 & -\beta \\
0 & 0 & \beta & 0
\end{array}
\right) x+\left(
\begin{array}{cccc}
0 & 0 & 0 & \gamma \\
0 & 0 & \delta & 0 \\
0 & -\delta & 0 & 0 \\
-\gamma & 0 & 0 & 0
\end{array}
\right) xu(t)=Ax+Bxu\left( t\right) .  \label{sso4}
\end{equation}
The coefficient matrices of the system belong to $so(4)$ and thus the system
evolves on the sphere, $S^{3},$ in $\mathbb{R}^{4}$. The constants, $\nu
,\beta ,\gamma $ and $\delta $ are positive and are related to the
inductances and capacitances of the elements of the circuits. Specifically,
we have,
\begin{equation*}
\nu =\frac{1}{\sqrt{L_{1}C_{2}}},\ \beta =\frac{1}{\sqrt{L_{3}C_{4}}},\
\gamma =\frac{1}{\sqrt{L_{1}C_{4}}},\ \delta =\frac{1}{\sqrt{L_{3}C_{2}}}.
\end{equation*}
Here $C_{1},C_{2}$ are the two capacitances and $L_{1},L_{3}$ are the two
inductances in the circuit (See \cite{c} for specific details). The state
vector $\mathbf{x}=\left( x_{1},x_{2},x_{3},x_{4}\right) ^{T}$ is defined as
$x_{1}=\sqrt{L_{1}}I_{1}$, $x_{2}=\sqrt{C_{2}}V_{2}$, $x_{3}=\sqrt{L_{3}}%
I_{3}$ and $x_{4}=\sqrt{C_{4}}V_{4}$. To this system we can associate two
systems whose unitary generators evolve on $SU(2)$ by using equations (\ref
{31}) - (\ref{33}) of section 2:
\begin{eqnarray}
\dot{U}_{1} &=&i\left( \frac{\nu +\beta }{2}\right) \sigma _{z}U_{1}-i\left(
\frac{\gamma +\delta }{2}\right) \sigma _{x}U_{1}u(t),\quad U_{1}\text{ in }%
SU(2)  \label{s1} \\
&=&A_{1}U_{1}+B_{1}U_{1}u(t)  \notag
\end{eqnarray}
and
\begin{eqnarray}
\dot{U}_{2} &=&i\left( \frac{\beta -\nu }{2}\right) \sigma _{z}U_{2}+i\left(
\frac{\gamma -\delta }{2}\right) \sigma _{x}U_{2}u(t),\quad U_{2}\text{\ in }%
SU(2)  \label{s2} \\
&=&A_{2}U_{2}+B_{2}U_{2}u(t).  \notag
\end{eqnarray}
Note that both systems are controlled by the \textit{same} control, $u(t)$.

The strategy to control the network is as follows. Supposed it is desired to
transfer the state from $(1,0,0,0)^{T}$ to a vector $y$ in $S^{3}$, then one
first represents $(1,0,0,0)^{T}$ and $y$ by the unit quaternions $1$ and $y$%
. The next step is to find a pair of unit quaternions $p$, $q$ such that $%
pq^{-1}=y$. To $p$ and $q$ there correspond matrices (denoted by $p$ and $q$
again) in $SU(2)$ (see section 2). We then try to find a $u(t)$ which will
prepare, simultaneously, $p$ for system (\ref{s1}) and $q$ for system (\ref
{s2}). In general there will be an infinite family $p$, $q$ such that $%
pq^{-1}=y$. To avail of this, we represent $p$ via equation (\ref{5}) with $%
\alpha ,\zeta ,\mu $ floating. This will then determine $q$. The parameters $%
(\alpha ,\zeta ,\mu )$ are then found by the requirement that the same $u(t)$
prepare both $p$ and $q$. The details of this strategy, of course, depend on
the specific sequence of piecewise constant controls which are used to
prepare a state for a given system on $SU(2)$. Equivalently, they depend on
the specific factorization of $SU(2)$ being employed.

We will now illustrate one such technique with $y=(0,0,1,0)^{T}$. The $SU(2)$
matrix corresponding to $y$ is $e^{i\frac{\pi }{2}\sigma _{y}}$. Thus, $%
pq^{-1}=e^{i\frac{\pi }{2}\sigma _{y}}$. It turns out that this strategy can
be implemented if any one of the following conditions on the constants of
the circuit holds:
\begin{equation}
\frac{\nu +\beta }{\beta -\nu }=2k+1=\frac{\gamma +\delta }{\delta -\gamma }%
,\quad k=1,2,\ldots  \label{cc1}
\end{equation}
The above conditions are, of course, artifices of the specific factorization
of $SU(2)$ that will be presently employed. Note that these conditions imply
that $C_{2}=C_{4}$ and that $\sqrt{\frac{L_{1}}{L_{3}}}=\frac{2k+2}{2k}$.

Representing $p$ as $S\left( \frac{\pi }{4},\frac{(2k+1)^{2}\pi }{4k(k+1)},-%
\frac{(2k+1)\pi }{4k(k+1)}\right) $ suffices. Indeed, one can now factorize $%
p$ as:
\begin{eqnarray}
p &=&\exp \left( \frac{(2k+1)\pi }{2(k+1)(\nu +\beta )}A_{1}\right) \exp
\left( \frac{(2k+1)\pi }{\sqrt{2}(\nu +\beta )}A_{1}+\frac{(2k+1)\pi }{\sqrt{%
2}(\delta +\gamma )}B_{1}\right)  \notag \\
&&\cdot \exp \left( \frac{(2k+1)\left( 6k+1\right) \pi }{2k\left( \nu +\beta
\right) }A_{1}\right) .  \label{dc1}
\end{eqnarray}

Since $q$ is determined by $p$ it follows that it can be factorized as:
\begin{eqnarray}
q &=&\exp \left( \frac{\pi }{2(k+1)(\beta -\nu )}A_{2}\right) \exp \left(
\frac{\pi }{\sqrt{2}(\beta -\nu )}A_{2}+\frac{\pi }{\sqrt{2}(\delta -\gamma )%
}B_{2}\right)  \notag \\
&&\cdot \exp \left( \frac{(6k+1)\pi }{2k\left( \beta -\nu \right) }%
A_{2}\right) .  \label{dc2}
\end{eqnarray}

Using equation (\ref{cc1}) it follows that the coefficients of $A_{1}$ match
those of $A_{2}$ and similarly the coefficients of $B_{1}$ match those of $%
B_{2}$. Since these coefficients represent the duration and power of the
pieces of the control $u(t)$, it follows that the same control, $u(t)$
prepares both $p$ and $q$ and thus achieves the desired state for the
circuit. Furthermore, the controls are indeed bang-bang with values $1$ or $%
0 $. This follows from equation (\ref{cc1}) which forces, $\frac{\delta
+\gamma }{\nu +\beta }=1$ (keeping in mind the relation of these constants
to the inductances and capacitances).

Several remarks are in order at this stage:

\begin{enumerate}
\item[i)]  $k=0$ was omitted from (\ref{cc1}) since it would be physically
unreasonable;\bigskip

\item[ii)]  Using similar ideas, it can be shown that under equation (\ref
{cc1}), an explicit pulse sequence can be found for state transfer from $%
(1,0,0,0)^{T}$ to any of the following states: $\pm (0,1,0,0)^{T},\pm
(0,0,1,0)^{T},\pm (0,0,0,1)^{T}$. Furthermore, this holds also when equation
(\ref{cc1}) is modified to $\left| \frac{\nu +\beta }{\beta -\nu }\right|
=2k+1=\left| \frac{\gamma +\delta }{\gamma -\delta }\right| ,\ k=1,2,\ldots $
For this it is useful to note that the Cayley-Klein representation (\ref{5}%
), need not necessarily be the polar coordinates of the entries. Indeed,
since $e^{i\pi }=e^{-i\pi }=-1$, one can begin with polar coordinates and
yet dispense with the restriction that $\alpha $ be in $[0,\frac{\pi }{2}]$.
To illustrate this, consider transferring the state from $(1,0,0,0)^{T}$ to $%
(0,0,0,1)^{T}.$ Suppose equation (\ref{cc1}) is replaced with
\begin{equation}
\frac{\nu +\beta }{\beta -\nu }=2k+1;\quad \frac{\gamma +\delta }{\delta
-\gamma }=-\left( 2k+1\right) ,\quad k=1,2,\ldots  \label{cc2}
\end{equation}
Now $pq^{-1}=e^{i\frac{\pi }{2}\sigma _{x}}.$ So if $p=S\left( \alpha ,\zeta
,\mu \right) ,$ then
\begin{equation}
q=S\left( \alpha -\frac{\pi }{2},\frac{\pi }{2}-\mu ,\frac{5\pi }{2}-\zeta
\right) .
\end{equation}
It suffices to choose $p=S\left( \frac{\pi }{4},\frac{(2k+1)\left(
5k+2\right) \pi }{4k\left( k+1\right) },\frac{(2k+1)\left( k-2\right) \pi }{%
4k\left( k+1\right) }\right) .$ Indeed, this $p$ can be factorized as
\begin{eqnarray}
p &=&\exp \left( \frac{(2k+1)3\pi }{2(k+1)(\nu +\beta )}A_{1}\right) \exp
\left( \frac{(2k+1)\pi }{\sqrt{2}(\nu +\beta )}A_{1}+\frac{(2k+1)\pi }{\sqrt{%
2}(\delta +\gamma )}B_{1}\right)  \notag \\
&&\cdot \exp \left( \frac{(2k+1)\left( 10k+2\right) \pi }{2k\left( \nu
+\beta \right) }A_{1}\right) .  \label{dc3}
\end{eqnarray}
Similarly, $q$ can be factorized as
\begin{eqnarray}
q &=&\exp \left( \frac{3\pi }{2(k+1)(\beta -\nu )}A_{2}\right) \exp \left(
\frac{\pi }{\sqrt{2}(\beta -\nu )}A_{2}+\frac{\pi }{\sqrt{2}(\gamma -\delta )%
}B_{2}\right)  \notag \\
&&\cdot \exp \left( \frac{(10k+2)\pi }{2k\left( \beta -\nu \right) }%
A_{2}\right) .  \label{dc4}
\end{eqnarray}
Thus the same control which prepares $p$ does likewise for $q.$ Once again,
this control is a bang-bang control with values 0 and 1;\bigskip

\item[iii)]  The state $(0,1,0,0)^{T}$ can be prepared by free evolution
without any conditions on the circuit. This is not evident from equation (%
\ref{sso4}), without the calculation of an exponential. On the other hand,
since $(0,1,0,0)^{T}$ is equivalent to $e^{i\frac{\pi }{2}\sigma _{z}},$ it
follows with minimal fuss upon passage to $SU(2),$ i.e., with no calculation
whatsoever. Indeed, the matrices $A_{1}$ and $A_{2}$ from equations (\ref{s1}%
) and (\ref{s2}) are (different) multiples of $i\sigma _{z}.$ Hence it
suffices to choose the targets $p$ and $q$ of systems (\ref{s1}) and (\ref
{s2}) as $p=e^{iL\sigma _{z}}$ and $q=e^{i\left( L-\frac{\pi }{2}\right)
\sigma _{z}},$ for some $L$ determined by
\begin{equation*}
\frac{L}{\nu +\beta }=\frac{L-\frac{\pi }{2}}{\beta -\nu }.
\end{equation*}
This $p$ and $q,$ and hence $(0,1,0,0)^{T},$ can obviously be prepared by
free evolution. Note, this conclusion did not even require an $su\left(
2\right) $ exponential. \bigskip

\item[iv)]  For final states other than those in ii), we believe the same
idea is viable. The resultant equations for the Cayley-Klein parameters of $%
p $ now are transcendental \cite{p}. Intuitively, it seems plausible that
these equations can be solved since $SO\left( 4\right) $ acts transitively
on the 3-sphere, $S^{3},$ with isotropy given by $SO\left( 3\right) .$ As $%
SO\left( 3\right) $ is nearly $SU(2),$ it seems reasonable to expect success
of the strategy of finding, parametrically, a suitable $p$ in $SU(2)$ to
ensure that the same $u(t)$ will prepare both $p$ and $q$;\bigskip

\item[v)]  The conditions imposed by equation (\ref{cc1}) or (\ref{cc2}) can
be met in practice. Nevertheless, it would be useful to achieve state
transfers without this restriction. For this, different factorizations of $%
SU(2)$ need to be developed;\bigskip

\item[vi)]  The factorizations in equations (\ref{dc1}), (\ref{dc2}), (\ref
{dc3}) and (\ref{dc4}) are \textbf{not} Euler factorizations. Even though
three factors appear in each of these expressions, it is clear that $A_{i}$
is not orthogonal to a linear combination of $A_{i}$ and $B_{i},$ $i=1,2.$
Thus, these factorizations are not Euler factorizations. Without passage to $%
SU(2),$ it seems formidable to find similar factorizations directly for $%
SO\left( 4\right) .$
\end{enumerate}

\section{Conclusions}

Lossless networks of the type studied in this paper are important in many
applications. Therefore, a constructive strategy for preparing desired
states in such circuits is interesting. In this paper the novel technique of
using factorizations of the special unitary group $SU(2)$ was shown to be a
viable mechanism for this issue. The key enabling factor is the rich
algebraic structure of $su(2).$ Finding similar formulae for $so(3)$ and $%
so(4)$ directly is harder. However, once a formula on $SU(2)$ has been found
- whether it be for exponentials, bang-bang controls etc., - it can be
transferred with ease to the orthogonal group. \textbf{This is the rationale
behind our method.}\ While the individual properties of the networks played
an important role in the success of the methodology, the basic idea of using
factorizations of $SU(2)$ is a useful complement to other methods for
dealing with systems evolving on the unitary and orthogonal matrices.
Indeed, for the treatment of systems with drift, techniques based on
decompositions of unitary groups appear to be more viable. It seems
reasonable to expect that similar methods should work for other systems
evolving on the orthogonal groups. For instance, $so(6)$ is isomorphic to $%
su(4)$. The latter Lie algebra plays an important role in quantum control.

\end{document}